\documentclass[numberedappendix]{emulateapj}

\newcommand{\himpc}{{\hbox {$~h^{-1}$}{\rm ~Mpc}}}
\newcommand{\tableskten}{&&&&\\[-6pt]}
\newcommand{\wmap}{{\it WMAP}~}
\newcommand{\jcap}{J. Cosmology Astropart. Phys.}

\slugcomment{Accepted to \textit{The Astrophysical Journal} 12/09/2007}

\shorttitle{Anisotropic Correlation Function of SDSS LRGs}
\shortauthors{Okumura et al.}

\begin{document}

\title{Large-Scale Anisotropic 
Correlation Function of SDSS Luminous Red Galaxies}

\author{
Teppei Okumura\altaffilmark{1}, 
Takahiko Matsubara\altaffilmark{1}, 
Daniel J. Eisenstein\altaffilmark{2}, 
Issha Kayo\altaffilmark{1}, \\
Chiaki Hikage\altaffilmark{1,3},
Alexander S. Szalay\altaffilmark{4}, 
\& Donald P. Schneider\altaffilmark{5}
}

\email{teppei@a.phys.nagoya-u.ac.jp}
\altaffiltext{1}
{Department of Physics, Nagoya University, Chikusa,
Nagoya, 464--8602, Japan}
\altaffiltext{2}
{Steward Observatory, University of Arizona, 933 N. Cherry Ave., 
Tucson, AZ 85121}
\altaffiltext{3}
{School of Physics and Astronomy, University of Nottingham, 
University Park, Nottingham, NG7 2RD, UK}
\altaffiltext{4}
{Department of Physics and Astronomy, The Johns Hopkins University,
Baltimore, MD 21218}
\altaffiltext{5}
{Department of Astronomy and Astrophysics, Pennsylvania State University, 
University Park, PA 16802}

\begin{abstract}
  We study the large-scale anisotropic two-point correlation function
  using 46,760 luminous red galaxies at redshifts 0.16 -- 0.47 from
  the Sloan Digital Sky Survey. We measure the correlation function as
  a function of separations parallel and perpendicular to the line of
  sight in order to take account of anisotropy of the large-scale
  structure in redshift space. We find a slight signal of baryonic
  features in the anisotropic correlation function, i.e., a ``baryon
  ridge'' corresponding to a baryon acoustic peak in the spherically
  averaged correlation function which has already been reported using
  the same sample. The baryon ridge has primarily a spherical
  structure with a known radius in comoving coordinates. It enables us
  to divide the redshift distortion effects into dynamical and
  geometrical components and provides further constraints on
  cosmological parameters, including the dark energy
  equation-of-state. With an assumption of a flat $\Lambda$ cosmology,
  we find the best-fit values of $\Omega_m = 0.218^{+0.047}_{-0.037}$
  and $\Omega_b = 0.047^{+0.016}_{-0.016}$ (68\% CL) when we use the
  overall shape of the anisotropic correlation function of
  $40<s<200\himpc$ including a scale of baryon acoustic
  oscillations. When an additional assumption of $\Omega_bh^2=0.024$
  is adopted, we obtain $\Omega_{\rm DE}=0.770^{+0.051}_{-0.040}$ and
  $w=-0.93^{+0.45}_{-0.35}$.  These constraints are estimated only
  from our data of the anisotropic correlation function, and they
  agree quite well with values both from the cosmic microwave
  background (CMB) anisotropies and from other complementary
  statistics using the LRG sample.  With the CMB prior from the 3 year
  \wmap  results, we give stronger constraints on those parameters.
\end{abstract}


\section{Introduction}

Recently, baryon acoustic oscillations have been observed in the
large-scale structure of the universe. These observations include an
analysis of the two-point correlation function (2PCF) of the Sloan
Digital Sky Survey (SDSS) Luminous Red Galaxy (LRG) spectroscopic
sample \citep[E05, hereafter]{E2005}, the power spectrum of the
Two-Degree Field (2dF) Galaxy Redshift Survey \citep{Cole2005} and the
SDSS LRG \citep{Hutsi2006a, Hutsi2006b, Tegmark2006, Percival2007a,
  Percival2007b}, and the angular power spectrum of the SDSS LRG
sample with photometric redshifts \citep{Padmanabhan2007,
  Blake2006}. There is a hint of acoustic oscillations in the SDSS
quasar sample \citep{Yahata2005}. These analyses have established the
ability of the baryon acoustic oscillations to constrain cosmological
parameters competitively and complementarily with the CMB
\citep[e.g.,][]{Spergel2007} and Type Ia supernovae \citep{Riess1998,
  Perlmutter1999}.

These previous analyses of the baryon oscillations, however, use
angle-averaged 2PCFs, $\xi(s)$, or angle-averaged power spectra,
$P(k)$, where $s$ and $k$ are the separation and wavenumber in
redshift space, respectively. A certain amount of information is lost
when anisotropies of structure are ignored. In their pioneering work,
\citet{AP1979} proposed that geometrical anisotropies in redshift
space of the high-z universe can be used as a probe of the
cosmological constant.  \citet{MS1996} and \citet{Ballinger1996}
pointed out that the anisotropy of the 2PCF and the power spectrum in
redshift surveys can constrain the dark energy components. Recently,
methods which directly use anisotropy of the baryon acoustic
oscillations have been theoretically developed for both the power
spectrum \citep{Hu2003, Seo2003, GB2005, Seo2007} and the 2PCF
\citep{M2004}. These approaches use functions of two variables:
separations parallel and perpendicular to the line of
sight. Observationally, estimations of such two-variable functions are
noisier than one-variable functions. The baryon acoustic signature in
the large-scale structure is weak even in one-variable statistics when
presently available samples of galaxies are used (see, e.g.,
E05). Therefore, methods that directly treat the anisotropy of the
baryon acoustic feature require very large survey volume
\citep{EHT1999,M2001}. \citet{Peacock2001} and \citet{Hawkins2003}
measured the 2PCF with two variables from the 2dFGRS
\citep{Colless2001} and detected the detailed signature of large-scale
coherent infall, and as a result were able to constrain a value of
$\beta \simeq \Omega_m^{0.6}/b$ which parameterizes linear redshift
distortions. However, they used information from scales much smaller
than the acoustic scale because of limited survey volume and did not
analyze the effect of geometrical distortion. On the other hand, the
correlation analyses of the 2dF QSO survey placed constraints on the
cosmological constant (Hoyle et al.~2002; Outram et al.~2004; da
\^Angela et al.~2005; see also Ross et al.~2007). Their analyses
still focused on smaller scales than the baryon acoustic scale.

In this paper we analyze the anisotropic 2PCF, including the baryon
acoustic peak in the large-scale structure and constrain relevant
cosmological parameters. We use a spectroscopic sample of the SDSS
LRG, which is the most useful sample for our purpose. Our analysis
differs from previous studies of baryon acoustic
oscillations in the LRG sample in that, owing to a theoretical development by \citet{M2004}, we take into consideration the fully
two-dimensional feature in the 2PCF to detect a geometrical distortion
effect; this is
the first cosmological application of the two-dimensional acoustic
peaks.

Before proceeding to the next section, we take note of nonlinearity on
large scales and the scale dependence of the galaxy biasing. The
nonlinear effects on the baryonic features that appear around
$100\himpc$ play an important role in taking account of percent-level
cosmology \citep[e.g.,][]{Meiksin1999, Seo2005, Jeong2006}. Recent
work also suggests that the scale-dependent biasing poses a serious
problem in analyzing galaxy surveys \citep[e.g.,][]{Blanton2006,
  Percival2007b, Smith2007, Coles2007,Sanchez2008}. However, for the
sake of simplicity, in this paper we consider only the large-scale
clustering and assume the biasing to be scale-independent and linear.



The structure of this paper is as follows. In \S~\ref{sec:sdss} we
describe the SDSS LRG sample used in our analysis. We then, in
\S~\ref{sec:2pcf}, measure the anisotropic 2PCF in redshift space and
estimate its covariance matrix. In \S~\ref{sec:theory} we outline the
cosmological parameter dependence on the modeled 2PCF, including
dynamical and geometrical distortions. Cosmological parameters are
constrained by the measured anisotropic 2PCF in
\S~\ref{sec:parameter}. In \S~\ref{sec:conclusion} our conclusions are
given.

\section{The SDSS LRG Sample}\label{sec:sdss} 

The SDSS \citep{Y2000, Stoughton2002}
 is an ongoing imaging and redshift survey which
uses a dedicated 2.5m telescope, a mosaic CCD camera, and two
fiber-fed double spectrographs \citep{Fukugita1996, Gunn1998,
  Gunn2006}. After image processing \citep{Lupton2001, Stoughton2002,
  Pier2003, Ivezic2004, Tucker2006} and calibration
\citep{Hogg2001, Smith2002}, the spectroscopic targets of LRGs are
selected from the imaging data according to the algorithm described by
\citet{E2001}. The tiling algorithm for the fibers is found in
\citet{Blanton2003a}. 

For our analysis, we use 46,760 LRGs over $3853 {\rm ~deg}^2$ in the
redshift range from 0.16 to 0.47, which is the same sample as the one
used in previous analyses of SDSS LRG clustering (Zehavi et al. 2005;
E05). The sky coverage is the same as lss\_sample14
\citep{Blanton2005} and is similar to that of the publicly available
SDSS Data Release 3 \citep{Abazajian2004}. The galaxies in the sample
have rest-frame {\it g}-band absolute magnitudes $-23.2<M_{g}<-21.2$
($H_0=100\mbox{~km~s}^{-1}\mbox{~Mpc}^{-1}$) with $K + E$
corrections of passively evolved galaxies to a fiducial redshift of
$0.3$ \citep{Blanton2003b}. The comoving number density of the sample
is close to constant out to $z=0.36$ (i.e., volume limited) because of
the narrow absolute magnitude cut, and drops thereafter due to the
flux limits \citep[see Fig. 1 of][]{Zehavi2005}. The radial and
angular selection functions, fiber collisions, and unobserved plates
are modeled using the method described in \citet{Zehavi2005}; E05
also provides the details of the sample.

\section{Anisotropic 2PCF of LRGs} \label{sec:2pcf}

\subsection{Measuring the LRG 2PCF}\label{sec:measure_2pcf}

\begin{figure}
\epsscale{.8}
\plotone{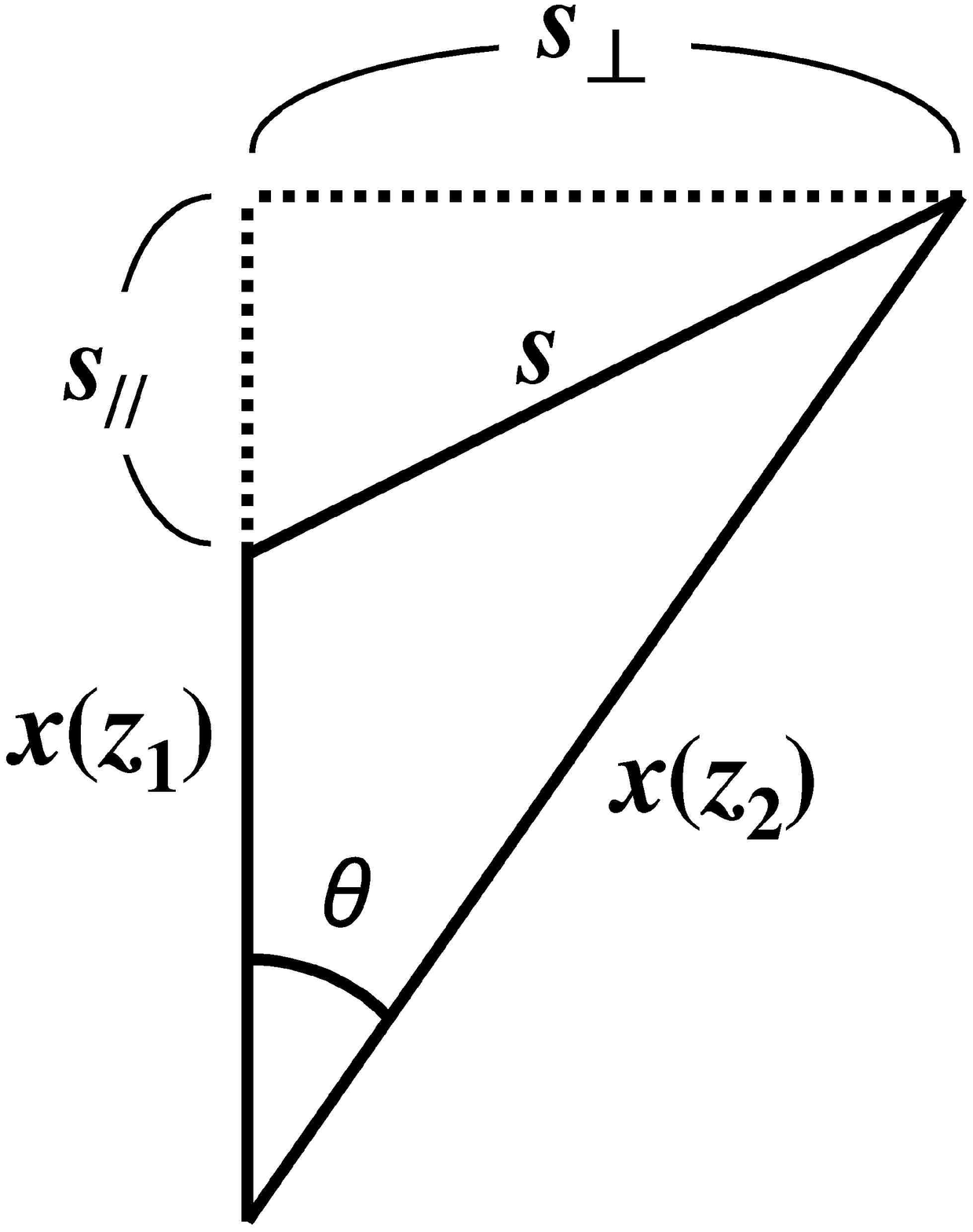}
\caption{Illustration of geometrical quantities used in this
  paper. }
\label{fig:coordinate_z}
\end{figure}

The 2PCF is measured by comparing the actual galaxy distribution to a
catalogue of randomly distributed points in the same region, according
to the selection function of the survey \citep{Peebles1980}. We count
the galaxy pairs in bins of comoving separation along and across the
line of sight, $s_{\parallel}$ and $s_{\perp}$ respectively, to
estimate the anisotropic 2PCF. Our notations of geometric quantities
are illustrated in Figure~\ref{fig:coordinate_z}.  First, the comoving
distances to every galaxy, $x(z)$, are calculated by assuming a flat
universe with $\Omega_m=0.3$ and $\Omega_{\Lambda}=0.7$, where
$\Omega_{\rm m}$ is the mass density parameter and $\Omega_\Lambda$ is
the cosmological constant parameter.  This flat universe is used only
as a mapping between the observed space and our analysis space: our
theoretical modeling also takes this mapping into account. The
purpose of the mapping is simply to avoid having to perform the
analysis in strongly distorted redshift space. For each galaxy
with redshift $z_1$ we define the separations $s_\parallel$ and
$s_\perp$ of other galaxies with redshift $z_2$ according to
Figure~\ref{fig:coordinate_z}:
\begin{eqnarray}
  s_\parallel &=& x(z_2) \cos\theta - x(z_1),
\\
  s_\perp &=& x(z_2) \sin\theta,
\end{eqnarray}
where $\theta$ is the apparent angle between the two galaxies from the
observer.  This definition of the line of sight is not as
standard as that of $(r_p,~\pi)$, such as was defined by
Davis \& Peebles (1983; see also Fisher et al. 1994).

Which galaxy of the pair is chosen as galaxy 1 is arbitrary in our
definition. Both galaxies are considered as galaxy 1, and the line of
sight is simply defined as the direction toward the galaxy
1. Therefore we count a pair of galaxies twice. As a result, there
appears a strong correlation between certain bins in
$(s_\perp,~s_\parallel)$ space; in particular, the bins of opposite
sign of $s_\parallel$ are strongly correlated. Those bins contain
almost identical pairs in a small-angle case, $\theta \ll 1$. However,
the angle $\theta$ in our sample is not always this small, so that those
bins contain different sets of pairs. Still the correlations between
those bins are strong, which is properly taken into account
in our parameter estimation below.

We compute the anisotropic 2PCF using the Landy-Szalay estimator
\citep{LS1993},
\begin{equation}
  \xi(s_{\perp},~s_{\parallel}) = \frac{DD-2DR+RR}{RR} \label{eq:ls},
\end{equation}
where $DD$, $RR$, and $DR$ are the normalized counts of galaxy-galaxy,
random-random, and galaxy-random pairs, respectively, in a particular
bin in the space of $(s_{\perp},~s_{\parallel})$. The random catalogue
contains about 30 times as many points as the real data, and the
random points are distributed according to the radial and angular
selection functions. The space of $(s_{\perp},~s_{\parallel})$ is
divided into rectangular cells with $\Delta s_{\perp},~\Delta
s_{\parallel} = 10\himpc$. Each galaxy of redshift $z$ is weighted by
$1/[1+n(z) P_w]$, where $n(z)$ is the comoving number density and
$P_w$ is the power spectrum at a typical scale \citep{Feldman1994}. We
adopt $ P_w = 40,000~h^{-3}\mbox{~Mpc}^3$, which is evaluated at the
baryon wiggle scale and is the same value as E05.  We have also tried
another value, $ P_w = 30,000h^{-3}\mbox{Mpc}^3$, adopted by
\citet{Tegmark2006}. We found that the value does not have a strong
effect on the result, as noted in \citet{Percival2007b}, because
the comoving number density of our sample is close to constant at
almost all scales.

\begin{figure}[thbp]
\epsscale{1.15}
\plotone{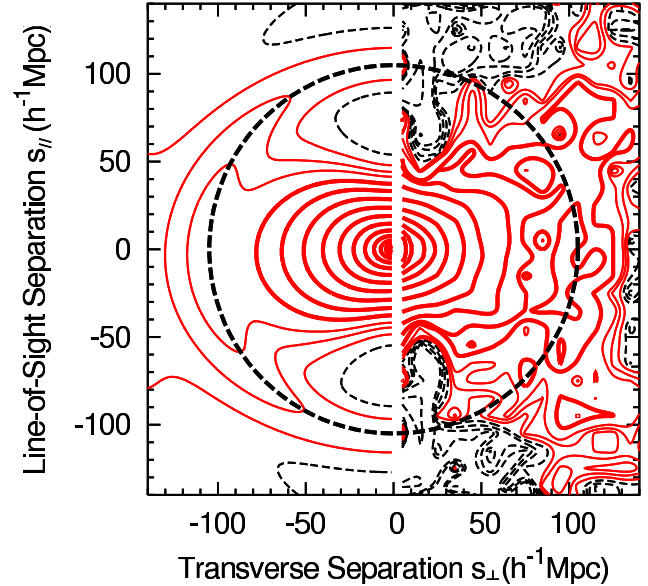}
\caption{Contour plots of the redshift-space 2PCF measured from the
  SDSS LRG sample ({\it right}) and the corresponding analytical
  formula derived by \citet{M2004} using a linear perturbation theory
  ({\it left}).  The dashed black lines show $\xi < -0.01$ increasing
  logarithmically with $0.25$ and $-0.01 \leq \xi < 0$ linearly with
  $0.0025$. The solid thin lines colored red show $0 \leq \xi <0.01$
  increasing linearly with $0.0025$ and the solid thick ones colored
  red $\xi \geq 0.01$ logarithmically with $0.25$. The baryonic
  feature slightly appears as ridge structures around the scale $s
  =(s_{\perp}^2+s_{\parallel}^2)^{1/2} \simeq 100\himpc$, and the
  dashed circle traces the peaks of the baryon ridges . For the
  theoretical predictions, we adopt the best-fit values assuming a
  flat cosmology, $\Omega_{\rm m}=0.218$, $\Omega_b=0.0473$,
  $h=0.702$, $\sigma_8=0.660$, $b=1.55$, while the fiducial values,
  $n_s=1$ and $w=-1$ are fixed. We also set the redshift at the origin
  to be $0.34$, which is typical in our LRG sample. }
\label{fig:xi_lrg_2d}
\end{figure}

The resulting redshift-space 2PCF for the observed LRGs is shown in
the right half of the plane in Figure~\ref{fig:xi_lrg_2d}.  The value
of $\xi(s_{\perp}, ~s_{\parallel})$ are given by contour lines. There
is an indication of the baryon ridges of radius about $100\himpc$,
which is a counterpart of the baryon peak in the one-dimensional 2PCF,
although the signal is not so strong. The anisotropy of the clustering is
obvious in this figure. When the separation is along the line of sight
($s_\perp \approx 0$), the clustering is elongated due to nonlinear
velocity dispersions of galaxies. On the other hand, the large-scale
clustering is squashed along the line of sight due to coherent infalls
toward over-dense regions \citep{Kaiser1987}. The latter effect is
often called the Kaiser's effect. A corresponding theoretical
prediction based on a linear perturbation theory derived by
\citet{M2000,M2004} is shown in the left half of the plane in
Figure~\ref{fig:xi_lrg_2d}.  Although the measured 2PCF is noisy, the
linear theory can account for the behavior of the 2PCF on large
scales. The nonlinear velocity distortions are not described by linear
dynamics, which should be removed in our linear analysis below. The
detailed comparison clearly needs statistical treatment, which we
explain below. When the anisotropic 2PCF is averaged over the angle,
the one-dimensional 2PCF $\xi(s)$ is obtained.

\subsection{Covariance Matrix} \label{sec:covariance} 

Because there are strong correlations between different bins of the
anisotropic 2PCF, it is necessary for statistically proper analyses to
estimate a covariance matrix. For this purpose, jackknife resampling
or bootstrap resampling \citep[e.g., see][]{Lupton1993} is often
adopted. In a cosmological context, however, cosmic variance plays a
critical role in estimates of cosmological parameters. It is uncertain
whether these methods can provide a reliable estimator of the cosmic
variance because they rely only on one observed sample. We first
tried to use the jackknife method and found that this approach
actually underestimates the covariance at all the scales, which leads
to the underestimation of the error bars for cosmological parameters
when we compare them with a more reliable method explained below.
A similar tendency is also seen by \citet{Pope2007}.

One of the best ways to estimate the covariance matrix including
cosmic variance is to use N-body simulations to generate many mock
catalogs from which the covariance matrix is calculated.  It is
necessary to generate a larger number of mock samples than the number
of data points of the statistics to be employed; otherwise, one would
improperly obtain a singular covariance matrix. However,
  because our analysis of the anisotropic 2PCF deals with several
  hundred data points,
it is computationally too expensive to produce a sufficient number of
independent realizations in our case.

\begin{figure}[tb]
\epsscale{1.05}
\plotone{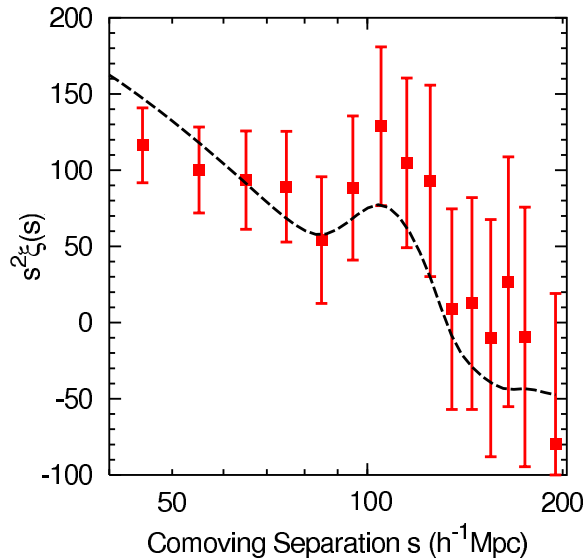}
\caption{Comparison of the 2PCFs times $s^2$ between the observed LRGs
  and the mock catalogs. The horizontal axis is logarithmic while the
  vertical axis is linear. The red points show the angle-averaged 2PCF of
  the LRGs and the error bars are from the mock catalogs.
  The dashed line shows the 2PCF averaged over the mock catalogs.  The
  baryon peak detected in these plots is obtained from the integration
  of baryon ridges in Fig. \ref{fig:xi_lrg_2d} over angular
  orientation.} \label{fig:xi_2lpt_1d}
\end{figure}

\begin{figure}[tb]
\epsscale{1.15}
\plotone{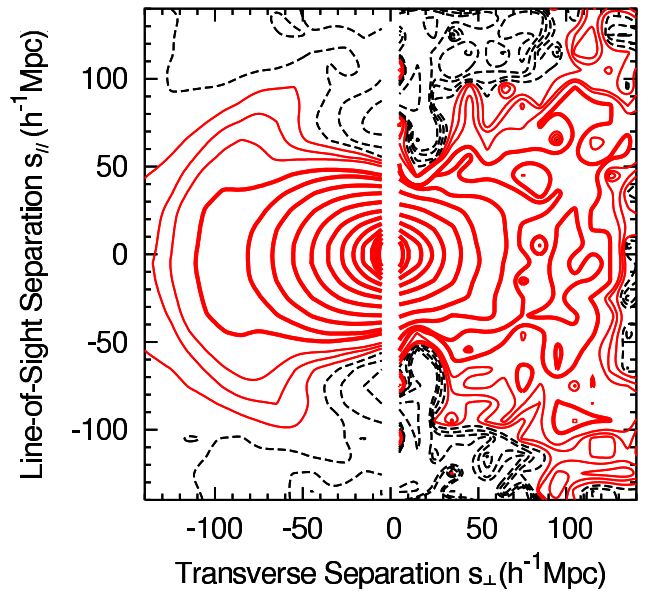} 
\caption{Anisotropic 2PCFs as functions of two variables, separations
  perpendicular and parallel to the line of sight. The right side
  shows the LRG 2PCF, which is the same as the right one of
  Fig. \ref{fig:xi_lrg_2d}. The left side shows the corresponding
  averaged 2PCF of our mock catalogs. The difference between 2PCFs in
  each mock and their average is used for construction of the
  covariance matrix. The 2PCF from our mock catalogs does not have
  large deviation from that of the observed 2PCFs, even for the
  quadrupole components. }
\label{fig:xi_2lpt_2d}
\end{figure}

In our analysis we adopt an alternative method, using a public
second-order Lagrangian perturbation theory (2LPT) code
\citep{Crocce2006}. As input ingredients, we adopt $\Omega _m=0.27$,
$\Omega_{\Lambda}=0.73$, $h=0.7$, $\sigma_8=0.8$, $256^3$ particles in
a cubic box of side $1600\himpc $, and a transfer function calculated
by CMBfast code \citep{Seljak1996} with $\Omega_b = 0.045$,
where $h$ is the Hubble parameter normalized by $100~{\rm
  km~s}^{-1}{\rm ~Mpc}^{-1}$, $\sigma_8$ is the rms. of the
fluctuations smoothed with a top-hat window function of radius
$R=8\himpc$, and $\Omega_b$ is the baryon density parameter. To
implant the galaxy biasing of the LRGs \citep{Zehavi2005}, we
empirically select particles with probability proportional to
$e^{\alpha\delta_m}$, where $\delta_m$ is the mass density
fluctuation at a position of each particle calculated by the
University of Washington HPCC's public SMOOTH
code.\footnote{At http://www-hpcc.astro.washington.edu/tools/smooth.html.}
We choose $\alpha =1.5$ so as to match the correlation amplitude of
the observed LRGs at scales larger than $40\himpc$. Then we trim the
mock catalogs from simulation boxes so as to have the same number
density and the same survey volume as the actual LRG sample.
Finally, we generate 2500 mock catalogs with independent initial
conditions, compute the 2PCF for each, and obtain a covariance matrix
by
\begin{equation}
  ({\it {\bf C}})_{ij}\equiv \mbox{Cov}(\xi_i,\xi_j)=
  \frac{1}{N-1}\sum^N_{l=1}(\xi^l_i-\bar{\xi}_i)(\xi^l_j-\bar{\xi}_j),
  \label{eq:cov}
\end{equation}
where $N=2,500$, $\xi^l_i$ represents the value of the 2PCF of $i$th
bin in $l$th realization, and $\bar{\xi}_i$ is the mean value of
$\xi_i^{l}$ over realizations. The average of 2PCFs from each mock
catalog agrees well with the observation within the $1~\sigma$ errors
for both one-dimensional (Fig.~\ref{fig:xi_2lpt_1d}) and
two-dimensional analyses (Fig.~\ref{fig:xi_2lpt_2d}). These
  mock catalogs are constructed solely to estimate errors of
  the measured 2PCF of LRGs, so the averaged 2PCF over the catalogs is
  not used for the following analyses.  Increasing $\alpha$ changes
the amplitude at the scales around the baryon peak to match the
observation, while it causes more discrepancy at the small
scales. This tendency is, however, consistent with the theoretical
prediction (see Figure 3 in E05).

\begin{figure}[tb]
\epsscale{1.15}
\plotone{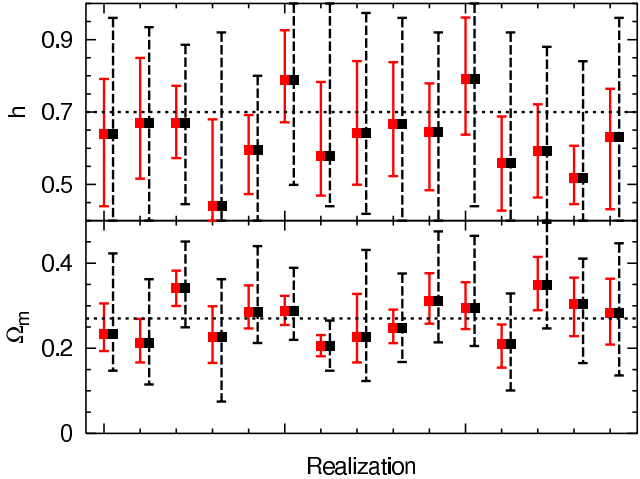}
\caption{Display of the reliability of the recovery of $\Omega_{\rm
    m}$ and $h$ from 15 realizations.  The horizontal axis shows the
  realization number, while the vertical axis shows the value of
  $\Omega_m$ and $h$ and the horizontal dashed lines show their input
  parameters.  Solid and dashed error bars shows the $68\%$ and $95\%$
  confidence levels, respectively.  Among these 15 realizations, there
  are 10 and 11 realizations including input values of $\Omega_m$ and
  $h$, respectively, at $68\%$ confidence intervals , and 14 and 15 at
  $95\%$ intervals.  }
\label{fig:cov_realization}
\end{figure}

We test the full covariance matrix obtained by the 2LPT method.
First, we randomly choose the 2PCFs of 30 realizations out of 2500. We
regard each of 30 2PCFs as an observed LRG sample, calculate the
$\chi^2$ statistics by the method described in \S \ref{sec:parameter},
constrain the input cosmological parameters, $\Omega_m$ and $h$, and
check how many realizations contain the input values at $68\%$ and
$95\%$ confidence level. If the amplitude of covariance is reasonable,
$68\%$ of all the contours of $68\%$ confidence levels should contain
the input parameters. The reason for using only 30 realizations is
that comparing against all the mock catalogs is computationally very
expensive; it requires that $\chi^2$ be calculated in seven
dimensional parameter space 2500 times. According to our statistics,
anisotropic 2PCF $\xi(s_{\perp},s_{\parallel})$, there are 22 and 28
realizations which contain the inputs at $68\%$ and $95\%$ confidence
levels for $\Omega_m$ in the total 30 realizations, while there are 17
and 28 for $h$. We thus conclude that our method of estimating the
covariance matrix is reasonable, and can be reliably applied for
parameter estimation in \S \ref{sec:parameter}. Figure
\ref{fig:cov_realization} shows the result of the test; we
choose to present only 15 realizations because displaying all 30
results makes the figure unclear.

\section{Theoretical Predictions} \label{sec:theory}

As a theoretical prediction for the anisotropic 2PCF, we adopt an
analytical formula of \citet{M2000,M2004} derived in a general
situation taking into account the wide-angle effect \citep{Szalay1998}
and the high-z distortion effect \citep{MS1996} in linear perturbation
theory. The necessary formula is given in \citet{M2004}.

The mean redshift of the LRG sample is about $0.34$ and the clustering
scale which we probe ranges up to $200\himpc$. The maximum angle between
two points from the observer ($\theta$ in
Fig.~\ref{fig:coordinate_z}) is $\approx 12\degr$. The
distant-observer approximation is not so accurate at $\gtrsim
10\degr$, and therefore the general formula give above,
which accurately includes the wide-angle effect,
is preferable for a precise analysis of the LRG sample.

The left side of Figure~\ref{fig:xi_lrg_2d} shows a prediction in
linear theory of the two-dimensional 2PCF in redshift space with the
central redshift of $z_1=0.34$. We adopt a flat cosmology with
$\Omega_m=0.218, \Omega_b=0.0473, h=0.702, \sigma_8=0.660$, 
$b=1.55$, $n_s=1$, and $w=-1$, where 
$w= p_{\rm DE}/\rho_{\rm DE}$ is the equation-of-state parameter for the
dark energy component and $b$ is the linear bias parameter. 
The first five values are our best-fit values for the
two-dimensional 2PCF of LRGs, as described in the following section, 
while the last two are the fiducial values.

Throughout this paper we assume a flat cosmology for simplicity. There
are seven cosmological parameters in our modeling: $\Omega_m$, $h$,
$\Omega_b$, $n_s$, $w$, $\sigma_8$, and $b$.  For the details of the
dependence of the anisotropic 2PCF on cosmological parameters, see
\citet{M2004}. In short, there are three kinds of physical
effects. The first one is the shape of the underlying mass power
spectrum, which is determined by the $\Omega_m$, $\Omega_b$, $h$, and
$n_s$. The second one is the dynamical distortion effect which is
generated by peculiar velocities of galaxies. Linear, coherent
velocities squash the apparent clustering along the line of sight,
while nonlinear, random velocities smear the clustering along the same
direction \citep{Kaiser1987, Hamilton1992}. The linear squashing
effect depends on the so-called redshift distortion factor, $\beta(z)
= f(z)/b(z)$, where $f(z) = d\ln D/d\ln a$ is the logarithmic
derivative of the linear growth rate $D(z)$ at redshift $z$, $a =
(1+z)^{-1}$ is the scale factor, and $b(z)$ is the linear bias factor
at redshift $z$.  The growth factor depends on $\Omega_m$ and
$w$. Since we assume a flat cosmology, the density parameter of dark
energy is given by $\Omega_{\rm DE} = 1 - \Omega_m$. However, the
parameter dependence on the growth factor is not so useful in
parameter estimation, because the overall amplitude of the power
spectrum characterized by $\sigma_8$ is a free parameter. Since the
parameter dependence of nonlinear velocity effect is not analytically
given, we do not use the nonlinear regime in the 2PCF. The third
effect is the geometric distortion, which depends on the Hubble
parameter $H(z)$, and angular diameter distance, $D_A(z)$, and
thus depends on $\Omega_m$, and $w$. The dependence of the geometric
distortion on $h$ vanishes in redshift surveys in which distances are
measured in units of $h^{-1}{\rm Mpc}$. The geometric distortion is
useful for constraining the dark-energy parameters, $\Omega_{\rm DE}$
and $w$.  The baryon ridges are isotropic in comoving space and their
anisotropy is primarily due to geometric distortion.

Finally, we comment on the galaxy biasing and the evolutionary effect.
Because we consider only the linear regime, we assume the biasing to
be scale-independent and linear. Although the bias parameter $b$ is
completely degenerate with $\sigma_8$ in the ordinary one-dimensional
2PCF, the two-dimensional 2PCF is able to solve this degeneracy
through the measurement of the redshift distortion parameter, $\beta$,
which depends on $b$. We treat $b$ and $\sigma_8$ as independent
parameters in the following analysis. One could choose $\beta$ or
$b\sigma_8$ as free parameters (which are more closely related to the
measurements), instead of $b$ or $\sigma_8$. The choice of the
independent parameters does not affect the following result.

In this paper we consider the measured 2PCF as a representative of
the function at a mean redshift $z_1 = 0.34$. We therefore simply
neglect the effects of evolution on clustering and biasing within the
sample. Strictly speaking, evolutionary effects are not negligible in
very large redshift surveys which have broad range of redshift
\citep[e.g.,][]{Yamamoto1999}. For example, the evolution has a
significant effect in the SDSS quasar sample \citep{Yahata2005}.
In our LRG sample, however, the redshift range is relatively
small and the signal-to-noise ratio of the measured 2PCF is not very
high.  Indeed, the evolutionary effect on the growth factor is about
less than 20\% from the survey edge to the mean redshift, but the
effects on the anisotropic 2PCF and cosmological parameters are
negligibly small compared to error levels. 

\section{Constraints on Cosmological Models}\label{sec:parameter} 

\subsection{Setup}

In this section we describe methods and results of constraining
cosmological parameters by the anisotropic 2PCF of the LRG sample.
We measure the goodness of fit, which shows how well assumed
  cosmological parameters fit a set of observational data, and the
measurements are given by the $\chi^2$ statistics, taking into
account the full covariance matrix. As described in the previous
section, we adopt seven free parameters $\Omega_m$
($=1-\Omega_{\rm DE}$), $\Omega_b$, $h$, $n_s$, $w$,
$\sigma_8$, and $b$, assuming flatness of the universe.

In comparing observational data with theory, we first compute the
theoretical 2PCF as a function of $(z_1, z_2,\theta)$ with a given set
of parameters.  Next we use the fiducial parameters of $\Omega_{\rm
  m}=0.3$ and $\Omega_{\Lambda}=0.7$ to convert redshifts into
comoving distances, which are the same values assumed for measuring
the distances of each galaxy in \S~\ref{sec:measure_2pcf}.  Therefore,
we compare the theoretical 2PCF with the observation in the same
comoving space and these fiducial parameters do not bias our results
of parameter estimations.

The theoretical formula based on the linear perturbation theory does
not reproduce nonlinear gravitational effects and nonlinear velocity
distortions such as finger-of-God effects. We therefore discard the
observed 2PCF at scales less than 40$\himpc$. We also do not use the
data along the line-of-sight, namely, $s_{\perp}<10\himpc$, because the
line-of-sight components of the 2PCF are noisy \citep{Bernstein1994}
and furthermore deviate from Kaiser's formula even on large scales
\citep{Scoccimarro2004}. Finally we perform the
analysis for the scale range of $40<s<200\himpc$. We also adopt a more
conservative range $60<s<160\himpc$ to check the systematic effects
beyond the linear theory. The numbers of bins in 2PCF are 574 for
$40<s<200\himpc$ and 330 for $60<s<160\himpc$. $\chi^2$ statistics
are then calculated as
\begin{equation}
  \chi^2({\bf\theta})=
  \sum_{i,j} \Delta\xi_{i}({\bf\theta})
  ({\bf C}^{-1})_{ij}\Delta\xi_j({\bf\theta}), \label{eq:chi2_general}
\end{equation}
where ${\bf\theta}$ is a set of cosmological parameters to be
constrained, $\Delta\xi_{i}({\bf\theta})$ denotes the difference
between the observed and theoretical 2PCFs in {\it i}th bin, and the
sum is over the number of bins. The most likely values for the
  cosmological parameters minimize the equation (\ref{eq:chi2_general}).
Finally, the likelihood function for the cosmological parameters,
${\cal L}$, is proportional to $\exp(-\chi^2/2)$ with an appropriate
normalization factor. Then, for example, the 68\% confidence
  interval becomes the region where $\int {\cal L}~d\theta = 0.68$ in
  the parameter space.

\subsection{$\sigma_8$-$b$ degeneracy}

\begin{figure}[htbp]
  \epsscale{1.15}
  \plotone{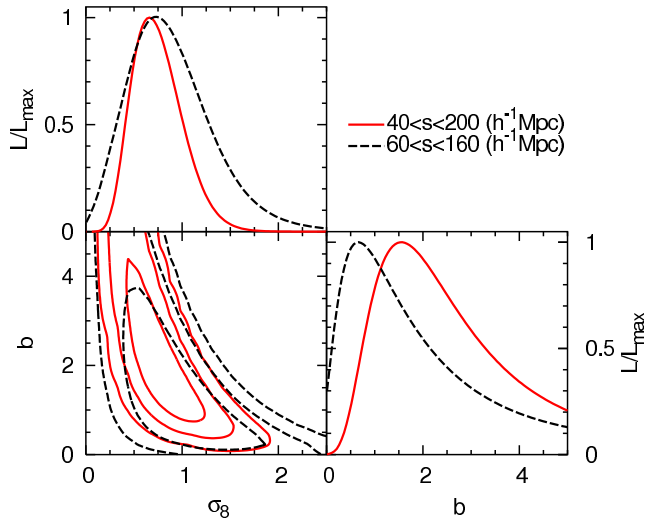}
 \caption{Likelihood contours for the parameters related to the
   clustering amplitude, $\sigma_8$ and $b$.  The diagonal panels
   represent the likelihood functions for the individual parameters,
   where $\Omega_m$, $\Omega_b$, $h$, and $n_s$ are marginalized
   over. The bottom left panel represents a two-parameter constraint,
   and each ellipse shows the 68\%, 95\%, and 99\% confidence levels
   from inward. The solid and dashed (colored red and black) contours
   are the fits for $40<s<200\himpc$ and $60<s<160\himpc$,
   respectively. The best-fit parameters are $\sigma_8=0.66$ and
   $b=1.55$ for $40<s<200\himpc$.  }
 \label{fig:l_s8_bi}
\end{figure}

First, we consider the behavior of the parameters related to the
clustering amplitude, $\sigma_8$ and $b$. Figure~\ref{fig:l_s8_bi}
plots their joint likelihood functions with contours representing
68\%, 95\% and 99\% confidence levels, where $w=-1$ is fixed and we
marginalize over the other four parameters $\Omega_m$, $\Omega_b$,
$h$, and $n_{\rm s}$. We find $\sigma_8=0.66^{+0.289}_{-0.216}$ and
$b=1.55^{+1.42}_{-0.75}$ (68\% confidence level) for the fit to
$40<s<200\himpc$ after also marginalizing them over each other. As
described in \S~\ref{sec:theory}, $b$ and $\sigma_8$ are strongly
coupled as the amplitude of the 2PCF is proportional to a product
$b\sigma_8$. The degeneracy is somewhat alleviated from anisotropy of
the 2PCF due to dynamical distortions which are dependent on $\beta$.
It is still difficult, however, to independently constrain these two
parameters without relying on other observations such as CMB, higher
order correlation analysis, etc. In this paper we mainly focus on
parameter constraints only from the 2PCF and consider the joint
analysis with the {\it Wilkinson Microwave Anisotropy Probe} (\wmap)
results only in \S~\ref{sec:wmap}.

Therefore, we always marginalize over both $\sigma_8$ and $b$ in the
following likelihood analysis. This marginalization corresponds to
mainly using the shape information in the 2PCF and discarding the
amplitude information.

We also note that the physical origins of the parameters $b$ and
$\sigma_8$ are not fully understood. The value of the bias parameter $b$
depends on unknown details concerning the formation of LRGs, and the value of
$\sigma_8$ depends on unknown details about the generation of density
fluctuations in the primordial universe. There is not any reliable
theory which robustly predicts the values of these parameters.

\subsection{Main results}

\begin{figure*}[htbp]
 \epsscale{1.}
\plotone{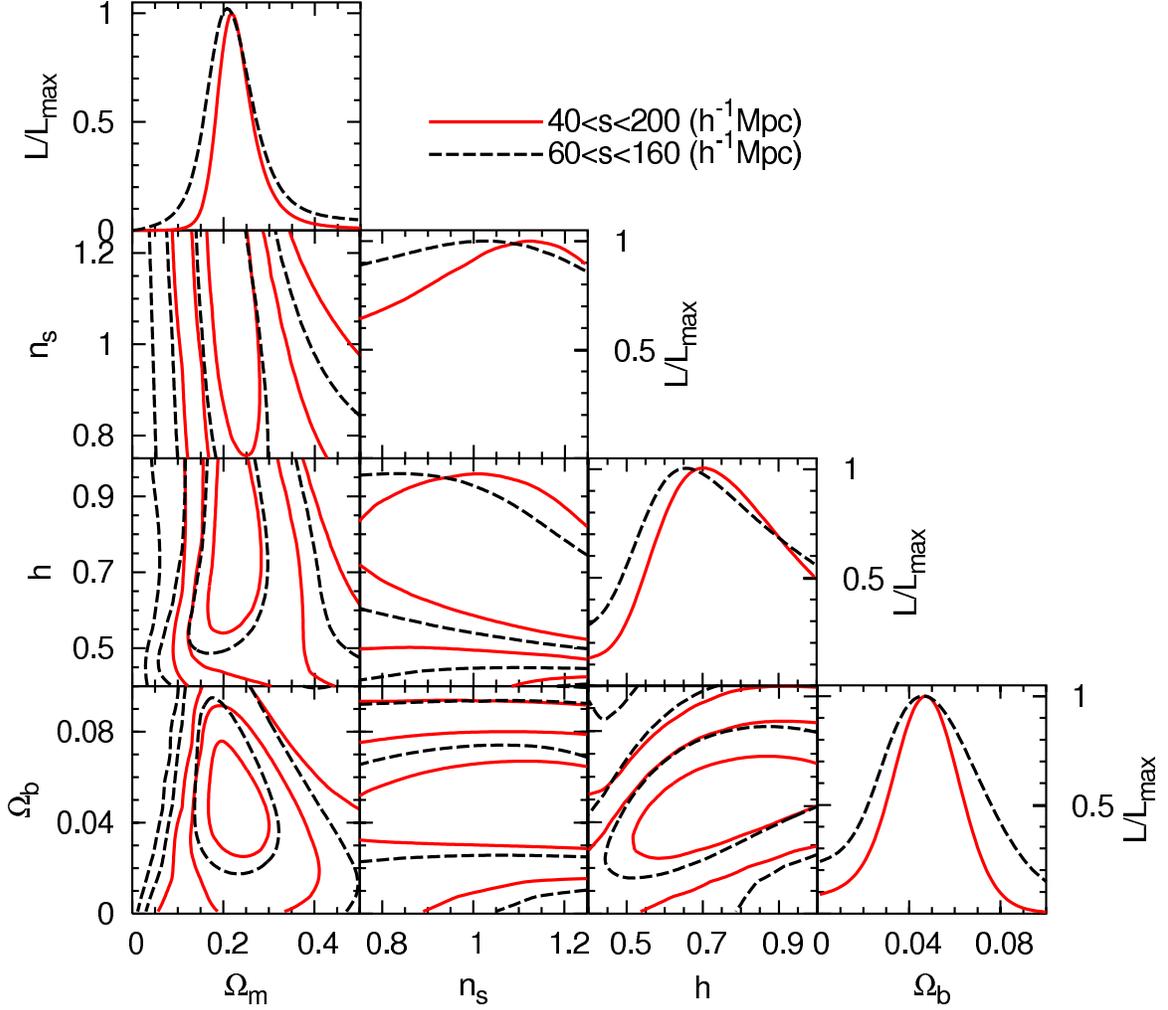} 
\caption{Likelihood contours for $(\Omega_m, \Omega_b,
  h,n_s)$, assuming a flat $\Lambda$CDM universe. The diagonal
  panels represent the likelihood functions for the four individual
  parameters with all the other parameters being marginalized over.
  The other panels show two-parameter constraints with the other
  parameters being marginalized, and each ellipse represents the
  constraint on the parameter space with 68\%, 95\%, and 99\% from
  inward. As in Fig. \ref{fig:l_s8_bi}, the solid (red) and dashed
  (black) contours are for $40<s<200\himpc$ and $60<s<160\himpc$,
  respectively. The best-fit parameters for $40<s<200\himpc$ are
  $\Omega_m=0.218$, $\Omega_b=0.0473$, $h=0.702$, and
  $n_s=1.122$ and the minimum value of $\chi^2$ is $\chi^2_{\rm
    min}=421.5$ with 568 dof. For $60<s<160\himpc$, the best-fit
  parameters are $\Omega_m=0.208$, $\Omega_b=0.0462$,
  $h=0.656$, and $n_s=1.030$ and $\chi^2_{\rm min}=216.6 $ with
  324 dof.}
\label{fig:l_m0_b0_h0}
\end{figure*}

We next focus on the four fundamental cosmological parameters
$\Omega_m$, $\Omega_b$, $h$, and $n_s$ after
marginalizing over $\sigma_8$ and $b$. Figure~\ref{fig:l_m0_b0_h0}
illustrates contour plots of the joint likelihood functions of two
parameters among the four, where $w$ is fixed at $-1$. 

The fits to $40<s<200\himpc$ give $\Omega_m=0.218^{+0.047}_{-0.037}$,
$\Omega_{\rm b}=0.0473^{+0.0157}_{-0.0160}$,
$h=0.702^{+0.187}_{-0.117}$, and $n_s=1.122^{+0.152}_{-0.183}$ ($68\%$
CL). The best-fit values and the errors of all the cosmological
parameters are listed in Table \ref{tb:bestfit}.  The best-fit values
of all of the four parameters for $40<s<200$ and $60<s<160\himpc$ are
consistent within the range of the $68\%$ error. This result suggests
that the systematic effects beyond the linear theory are small. The
accuracy of the constraints on $\Omega_m$ and $\Omega_b$ increases by
nearly a factor of 2 while the uncertainty of $h$ is improved only
marginally when the wider dynamical range is used. We obtain a
relatively strong constraint on $\Omega_m$ without fixing the value of
$\Omega_b$ because the constraint from the linear squashing effect in
the anisotropic 2PCF is stronger than in the spherically averaged
2PCF. In addition, the best-fit value of $\Omega_m$ is in quite good
agreement with an independent analysis of the LRGs by
\citet{Percival2007b} using a power spectrum that yields $\Omega_m
=0.22\pm 0.04$ with a dynamical range of $0.01<k<0.06h{\rm
  Mpc}^{-1}$. Our result also agrees with those from the CMB angular
power spectra from \wmap \citep{Spergel2007}.

\begin{figure}[tb]
\plotone{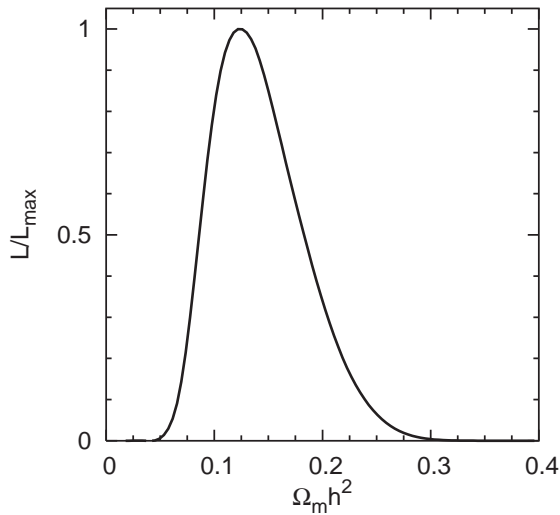}
\epsscale{1.}
\caption{Likelihood function for a parameter $\Omega_m h^2$,
  fitted for the scale range of $40<s<200\himpc$. The parameters $h$,
  $\sigma_8$, and $b$ are marginalized over, and $\Omega_b
  h^2=0.024$ and $n_s=0.98$ is fixed.  For the best-fit
  parameter, $\chi_{\rm min}^2 = 422.2$ with 570 dof.}
\label{fig:l_mh_h0}
\end{figure}

We note that the best-fit value of $\Omega_m$ is smaller (but not
significantly so) than the result of E05, $\Omega_m=0.273\pm
0.025$. This is not surprising because the analytical methods differ
from each other. The analysis in E05 is based on the spherically
averaged 2PCF and uses information from smaller scales ($s<40\himpc$).
They fix the baryon density parameter as $\Omega_b h^2=0.024$.  We
obtain a constraint of $\Omega_mh^2 = 0.123^{+0.048}_{-0.032}$ from
our data when the parameter $\Omega_b h^2=0.024$ is fixed, as listed
in Table \ref{tb:bestfit} (see also Fig.~\ref{fig:l_mh_h0}). This
result is consistent with the E05 result, $\Omega_mh^2=0.130 \pm
0.011$.
\begin{deluxetable*}{c|ccccc}
  \tablecaption{\label{tb:bestfit}Summary of constraints on cosmological parameters}
    \startdata
      \hline \hline \tableskten    
       & \multicolumn{2}{c}{LRG only} & LRG($40<s<200$) &  & \\ 
      Parameter &$40<s<200$ & $60<s<160$ & +WMAP3      & Marginalized  & Fixed \\ 
      \hline \hline\tableskten\tableskten
      $\Omega_{\rm m}$ & $0.218^{+0.047}_{-0.037}$ & $0.208^{+0.069}_{-0.055}$ & $0.240^{+0.019}_{-0.025}$ 
      & $\Omega_{\rm b},h,n_{\rm s},\sigma_8,b$ & $w$\\ \tableskten
      $\Omega_{\rm b}$ & $0.0473^{+0.0157}_{-0.0160}$ & $0.0462^{+0.0253}_{-0.0208}$ & $0.0414^{+0.0023}_{-0.0024}$ 
      & $\Omega_{\rm m},h,n_{\rm s},\sigma_8,b$ & $w$\\\tableskten
      $h$             & $0.702^{+0.187}_{-0.117}$  & $ 0.656^{+0.220}_{-0.120}$   & $0.718^{+0.023}_{-0.020}$ 
      & $\Omega_{\rm m},\Omega_{\rm b},n_{\rm s},\sigma_8,b$ & $w$\\\tableskten
      $n_{\rm s}$      & $1.122^{+0.152}_{-0.183}$  & $1.030^{+0.144}_{-0.189}$    & $0.947^{+0.016}_{-0.015}$ 
      & $\Omega_{\rm m},\Omega_{\rm b},h,\sigma_8,b$ & $w$\\\tableskten
      $\sigma_8$      & $ 0.660^{+0.289}_{-0.216}$& $0.728^{+0.471}_{-0.359}$ &$0.736^{+0.050}_{-0.062}$ 
      & $\Omega_{\rm m},\Omega_{\rm b},h,n_{\rm s},b$ & $w$\\\tableskten
      $\Omega_{\rm DE}$ & $0.770^{+0.051}_{-0.040}$ & $0.786^{+0.060}_{-0.061}$ & $0.772^{+0.024}_{-0.033}$
      & $h,n_{\rm s},w,\sigma_8,b$ & $\Omega_{\rm b}h^2$\\ 
      \tableskten\tableskten
      $w$        & $-0.93^{+0.45}_{-0.35}$ & $-1.07^{+0.49}_{-0.46}$ 
      & $-0.97^{+0.12}_{-0.11}$ & $\Omega_{\rm DE},h,n_{\rm s},\sigma_8,b$ & $\Omega_{\rm b}h^2$\\ 
      \tableskten\tableskten\hline\tableskten\tableskten
      $\Omega_{\rm m} $ & $0.235^{+0.040}_{-0.037}$ & ---
      & ---&  $h,\sigma_8,b$ & $\Omega_{\rm b}h^2$, $n_{\rm s}$, $w$ \\
      \tableskten\tableskten
      $\Omega_{\rm m} h^2$ & $0.123^{+0.048}_{-0.032}$ & ---& ---
      &$h,\sigma_8,b$ & $\Omega_{\rm b}h^2$, $n_{\rm s}$, $w$
      \enddata
      \tablecomments{The comoving distance $s$ is in units of
        $\himpc$. We assume $\Omega_{\rm b}h^2=0.024$, $w=-1$, and
        $n_{\rm s}=0.98$ when they are fixed.}
\end{deluxetable*}

All the constraints obtained above are more conservative than those in
other work using the baryon acoustic oscillations because we neglect
the small-scale data and all CMB information. 
For example, we obtain worse constraints on $h$ than those on
$\Omega_m$. A possible reason for this is that discarding the
small-scale data makes the constraints on $\Omega_mh^2$
degenerate along the direction of constant acoustic scales (see Fig. 8 of E05) and this direction is fairly parallel to the lines of
constant $\Omega_m$ but intersects with the lines of constant
$h$ (Fig. 10 of E05).

\subsection{Dark energy constraint}

\begin{figure}[tb]
  \epsscale{1.15}
  \plotone{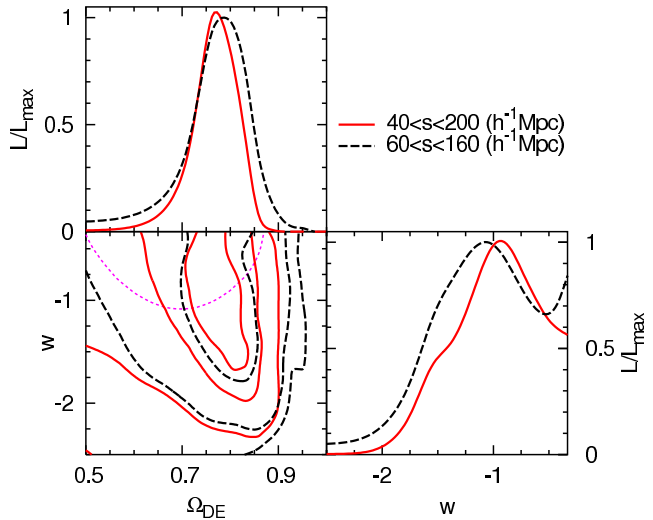} 
\caption{Dark energy constraints on $\Omega_{\rm DE}$ and $w$ under
  the assumption of a flat universe and $\Omega_bh^2=0.024$. As in
  Fig. \ref{fig:l_m0_b0_h0}, the bottom left panel represents a
  two-parameter constraint, $\Omega_{\rm DE}$-$w$, with $h$, $n_s$,
  $\sigma_8$ and $b$ marginalized over, and each ellipse represents
  the constraints on the parameter space with 68\%, 95\% and 99\% from
  the inside. The solid (red) and dashed (black) contours are for
  $40<s<200$ and $60<s<160\himpc$, respectively. The two diagonal
  panels represent likelihood functions with all the other four
  parameters marginalized over. The best-fit parameters are
  $\Omega_{\rm DE}=0.770$, $w=-0.93$, where $\chi_{\rm min}^2=421.2$
  with $568$ dof for $40<s<200\himpc$, while $\Omega_{\rm DE}=0.786$,
  $w=-1.07$ and $\chi_{\rm min}^2=216.4$ with $324$ dof.~for
  $60<s<160\himpc$. The dotted (magenta) contour in the bottom left
  panel shows the 68\% CL when we smooth the oscillatory part of the
  transfer function and use it for the fit of $40<s<200\himpc$. }
\label{fig:l_l0_w0}
\end{figure}

Constraining the dark energy is one of the most interesting
applications of the anisotropic 2PCF. As described in
\S~\ref{sec:theory}, we use the information from not only the overall
shape of the 2PCF but also the geometrical distortion with baryon
ridges to constrain the dark energy component. As \citet{M2002}
indicated, the LRG sample is one of the best samples for probing the
feature of dark energy among current redshift surveys. However, it is
still difficult for the relatively low-$z$ survey to constrain not
only the value of $w$ but also its evolution. In this subsection, we
therefore assume $w$ to be a constant not necessarily equal to
$-1$ and fix $\Omega_bh^2 =0.024$ because the baryon density is
highly constrained by the analysis of the \wmap data
\citep{Spergel2007} and big bang nucleosynthesis
\citep{Burles2001}. In Figure \ref{fig:l_l0_w0} we plot the joint
likelihood functions of $\Omega_{\rm DE}$ and $w$ and obtain
$\Omega_{\rm DE}=0.770^{+0.051}_{-0.040}$ and
$w=-0.93^{+0.45}_{-0.35}$ for the fit to $40<s<200\himpc$. All the
other parameters but $\Omega_bh^2$ are marginalized over. Our
constraints on dark energy parameters are listed in Table
\ref{tb:bestfit}.

We also plot the likelihood function in the bottom left panel of
Figure \ref{fig:l_l0_w0} where the no-wiggle power spectrum
\citep{EH1998} is used to calculate the analytical formula for the
anisotropic 2PCF. Because the dark energy parameter is constrained
only from the overall shape of the 2PCF in this way and degenerates
with the other parameters without information from the acoustic scale,
we obtain a poorer fit to the data.  Therefore, the overall shape is
not a dominant effect in constraining the dark energy parameter, and
the baryon ridges contribute as well.

\subsection{Combining with the \wmap results}
\label{sec:wmap}
So far we have focused on parameter constraints using the LRG data only,
which are very useful to check its result independently, while the
obtained constraints are inevitably weaker than those we would obtain
when other data sets are combined. In this subsection we consider the
additional constraints using the CMB prior
from the 3 year \wmap data \citep{Spergel2007}.

We consider two Markov chain Monte Carlo results of the \wmap data,
$w=-1$ and constant $w$ cosmologies \citep{Tegmark2006}. We find
$\Omega_m=0.240^{+0.019}_{-0.025}$, $\Omega_{\rm
  b}=0.0414^{+0.0023}_{-0.0024}$, $h=0.718^{+0.023}_{-0.020}$, $n_{\rm
  s}=0.947^{+0.016}_{-0.015}$, $\sigma_8=0.736^{+0.050}_{-0.062}$, and
$b=2.15^{+0.28}_{-0.36}$ from the former chain, while $\Omega_{\rm
  DE}=0.772^{+0.024}_{-0.033}$ and $w=-0.97^{+0.12}_{-0.11}$ from the
latter.  These constraints are also summarized in Table
\ref{tb:bestfit} and are in very good agreement with the previous
studies for the joint constraints of the \wmap observation with the
large-scale structure \citep[e.g.,][]{Tegmark2006,
  Spergel2007}. Although a pure LRG analysis cannot tightly constrain
$\Omega_mh^2$ or $h$ because of the limited range of
separations, they are significantly improved by the prior on the CMB
acoustic scale.

\section{Conclusions}
\label{sec:conclusion}

We have presented the 2PCF in redshift-space for the SDSS Luminous Red
Galaxy sample considering the anisotropy in 2D redshift space. In
particular, we have focused on the distorted features of the 2PCF in
redshift-space from both peculiar velocities of galaxies and
geometrical effect. The distorted features of the Kaiser and
finger-of-God effects were clearly detected. The baryon ridges, which
are the baryonic acoustic features in the anisotropic 2PCF, are a
nearly spherical object in comoving space. We found indications of
baryon ridges in the measured 2PCF. Beyond qualitative comparison
between data and theory, evaluation of the covariance matrix is needed
for cosmological parameter estimation. We constructed the matrix by
generating mock samples using the second-order Lagrangian perturbation
theory with an artificial biasing scheme. We have constrained the
cosmological parameters by comparing the observed 2PCF with linear
theory.

We have obtained constraints on fundamental cosmological parameters,
$\Omega_m = 0.218^{+0.047}_{-0.037}$, $\Omega_b =
0.0473^{+0.0157}_{-0.0160}$, $h = 0.702^{+0.187}_{-0.117}$ , and
$n_s=1.122^{+0.152}_{-0.183}$ when we have used the data of
$40<s<200\himpc$.  The constraint on $\Omega_m$ was better
mainly because of the clear detection of the Kaiser effect, which
directly depends on $\Omega_m$ through $\beta$. We have also
obtained the constraints on the dark energy as $\Omega_{\rm
  DE}=0.770^{+0.051}_{-0.040} $ and $w=-0.93^{+0.45}_{-0.35}$ when we
fix $\Omega_bh^2=0.024$ and the other parameters, $h$, $n_{\rm
  s}$, $\sigma_8$, and $b$ are marginalized over.  These constraints
are mainly due to the overall shape of the anisotropic 2PCF and the
information from geometrical distortions including the scale of the
baryon ridge.  We have demonstrated that a pure LRG analysis can
constrain $w$ by considering the anisotropy of the structure
accurately.  As for the parameters related to the clustering
amplitude, we have obtained $\sigma_8=0.66^{+0.289}_{-0.216}$ and
$b=1.55^{+1.42}_{-0.75}$.  While these two parameters are strongly
coupled, the degeneracy was alleviated from anisotropy of the 2PCF
through the redshift distortion factor, $\beta$.  In addition,
stronger constraints on the cosmological parameters above were
obtained by the CMB prior from the 3 year \wmap results.  All the
constraints summarized above agree with the previous studies in
literature.  

The current analysis can be improved by considering two issues below.
The first issue is theoretical improvement accounting for the
nonlinearity of the gravitational evolution, the redshift distortions
and the galaxy biasing. Although the baryonic signature emerges on
very large scales, the width of the baryon peak in the 2PCF is an
order of $10~$Mpc. The nonlinearities nontrivially affect such a 
feature. In fact, such effects have already been investigated using
$N$-body simulations, higher-order perturbation theories, and
renormalization perturbation theory \citep{Meiksin1999, Seo2005,
  Springel2005, Jeong2006, Crocce2007, Nishimichi2008, M2007}. The
degradation of the acoustic signature was well modeled; it was shown
that the acoustic peak of the linear density field in the 2PCF can be
reconstructed \citep{E2007a, E2007b}. In addition, the overall shape
of the redshift-space 2PCF is also affected by nonlinear dynamics
\citep{Scoccimarro2004}. According to his result, the redshift-space
2PCF for pairs parallel to the line of sight in a random Gaussian
field deviates from the prediction of standard linear theory even on
fairly large scales. In this work we do not use the data along the
line of sight; however, we shall include these issues in the future
analysis. The theoretical and numerical studies also suggest that the
biasing is potentially scale dependent even on large scales
\citep[e.g.,][]{Schulz2006, Smith2007, Coles2007}, which poses a serious
problem for estimating cosmological parameters from galaxy surveys
\citep{Blanton2006, Percival2007b, Sanchez2008}.

The second issue is the calculation of the covariance matrix for the
measured 2PCF. As described in \S~\ref{sec:covariance}, we have
constructed a covariance matrix by the second-order perturbation
theory. Using N-body simulations which fully include nonlinearity
provides better estimation of the covariance matrix. However, 
this approach is too computationally expensive to produce 
a large number of independent realizations.
The approximation by the second-order perturbation theory is valid
on scales which we consider in this work. However, in order to
utilize the information at smaller scales for more accurate
cosmological parameter estimation, we must estimate the
nonlinearities more accurately using $N$-body simulations or more
sophisticated methods such as a halo occupation model from the
second-order Lagrangian perturbation theory
\citep[e.g.,][]{Scoccimarro2002}. 

The most important point of our analysis is that we directly include
anisotropies of the structure. The baryonic features enable to divide
the effect of the redshift distortions into dynamical and geometrical
components. The anisotropy due to the geometric distortion, in
particular, contributes to better estimation of the equation-of-state
parameter for the dark energy. Various methods using the scale of the
oscillations as a standard ruler have been considered for both the
power spectrum and the 2PCF \citep{EH1998, BG2003, Hu2003, Seo2003,
  M2004, GB2005}. This work is the first application of the anisotropy
in the 2PCF with baryon acoustic features to observational data, which
was proposed by \citet{M2004}. Direct measurement of the
  growth function from the Kaiser's effect and the two-dimensional
  acoustic scales which depend on $D_A(z)$ and $H(z)$ is also an
  attractive challenge of the analysis using the anisotropic 2PCF,
  so these topics will be definitely pursued in future work with an
  improved LRG sample.

The baryonic signature from the redshift range of SDSS LRGs is not
strong, because their number density is relatively small and nonlinear
effects weaken the baryonic feature. There are many plans for
constraining the dark energy by future wide-field, deep galaxy
surveys: the Fiber Multiobject Spectrograph
\citep[FMOS;][]{Kimura2003}, Wide-Field Multiobject Spectrograph
\citep[WFMOS; ][]{GB2005, Bassett2005}, Baryon Oscillation Probe
\citep[BOP; ][]{Glazebrook2005}, 
and the Hobby-Eberly Dark Energy Experiment
\citep[HETDEX; ][]{Hill2004}, and so on. When the baryonic signature is
detected with high accuracy from future redshift surveys, the analysis
of the anisotropic 2PCF as in this work will be an important
ingredient for stringently constraining properties of the dark energy.

\acknowledgments 

We acknowledge helpful discussions with Naoshi Sugiyama, Roman
Scoccimarro, and Kazuhiro Yahata. We also thank the anonymous referee
for useful comments. This work is supported in part by Grand-in-Aid
for Scientific Research on Priority Areas, No. 467, Probing the Dark
Energy through an Extremely Wide and Deep Survey with the Subaru
Telescope, and by the Mitsubishi Foundation. T.~M. acknowledges
support from the Ministry of Education, Culture, Sports, and
Technology (MEXT), and Grant-in-Aid for Scientific Research
(No. 18540260). I.~K. acknowledges support from the MEXT, and a
Grant-in-Aid for Encouragement of Young Scientists (No. 17740139).
C.~H. acknowledges support from a JSPS (Japan Society for the
Promotion of Science) fellowship. C.~H. also acknowledges the support
from the Particle Physics and Astronomy Research Council grant 
PP/C501692/1. Numerical calculations are performed by a parallel
computing system at Nagoya University. Funding for the SDSS and
SDSS-II has been provided by the Alfred P. Sloan Foundation, the
Participating Institutions, the National Aeronautics and Space
Administration, the Japanese Monbukagakusho, the Max Planck Society,
and the Higher Education Funding Council for England. The SDSS Web
Site is http://www.sdss.org/.

The SDSS is managed by the Astrophysical Research Consortium for the
Participating Institutions. The Participating Institutions are the
American Museum of Natural History, Astrophysical Institute Potsdam,
University of Basel, University of Cambridge, Case Western Reserve
University, University of Chicago, Drexel University, Fermilab, the
Institute for Advanced Study, the Japan Participation Group, Johns
Hopkins University, the Joint Institute for Nuclear Astrophysics, the
Kavli Institute for Particle Astrophysics and Cosmology, the Korean
Scientist Group, the Chinese Academy of Sciences (LAMOST), Los Alamos
National Laboratory, the Max Planck Institute for Astronomy (MPIA),
the Max-Planck-Institute for Astrophysics (MPA), New Mexico State
University, Ohio State University, University of Pittsburgh,
University of Portsmouth, Princeton University, the United States
Naval Observatory, and the University of Washington.


\end{document}